\documentclass[11pt]{article}
\usepackage{cite}
\usepackage{amsmath,amsfonts,amssymb}
\usepackage{graphicx,color}
\usepackage{epsfig,epsf}
\usepackage{bm}
\usepackage{dcolumn,subfig}
\usepackage{array}

\usepackage{relsize}
\def\Babar{{\mbox{\slshape B\kern-0.1em{\smaller A}\kern-0.1em B\kern-0.1em{\smaller A\kern-0.2em R}}}}

\usepackage[bookmarks, breaklinks, colorlinks,urlcolor=black, citecolor=red, 
linkcolor=blue]{hyperref}

 \usepackage{color}

 \usepackage[normalem]{ulem}

 \definecolor{darkgreen}{cmyk}{1,0,1,0.4}
 
 \definecolor{pink}{cmyk}{0.4,1,0.3,0}

\def\com2#1{\textcolor{red}{\it{#1}}}

\def\bar {\overline}

\def\bra {\langle}
\def\ket {\rangle}

\def\beq{\begin{equation}}
\def\eeq{\end{equation}}
\def\bea{\begin{eqnarray}}
\def\eea{\end{eqnarray}}
\def\barr{\begin{array}}
\def\earr{\end{array}}
\def\nn {\nonumber}

\def \gsrr {g^S_{RR}}
\def \gsrl {g^S_{RL}}
\def \gslr {g^S_{LR}}
\def \gsll {g^S_{LL}}
\def \gvrr {g^V_{RR}}
\def \gvrl {g^V_{RL}}
\def \gvlr {g^V_{LR}}
\def \gvll {g^V_{LL}}

\def\gev{\ensuremath{\mathrm{Ge\kern -0.1em V}}}

\begin{document}

\renewcommand*{\thefootnote}{\fnsymbol{footnote}}

\begin{center}
 {\Large\bf{New physics with the lepton flavor violating decay $\bm{\tau\to 3\mu}$}}\\[5mm]
{\bf Zaineb Calcuttawala} $^{a,}$\footnote{zaineb.calcuttawala@gmail.com}, 
{\bf Anirban Kundu} $^{a,}$\footnote{anirban.kundu.cu@gmail.com}, \\
{\bf Soumitra Nandi} $^{b,}$\footnote{soumitra.nandi@gmail.com}, and  
{\bf Sunando Kumar Patra} $^{b,c,}$\footnote{sunando.patra@gmail.com}\\[3mm]
$^a$ Department of Physics, University of Calcutta, \\
92 Acharya Prafulla Chandra Road, Kolkata 700009, India\\ 
$^b$  Department of Physics, Indian Institute of Technology, Guwahati 781039, India\\
$^c$  Department of Physics, Indian Institute of Technology, Kanpur 208016, India
 \\ 
 \today
 \end{center}


\begin{abstract}

Lepton flavour violating (LFV) processes are a smoking gun signal of new physics (NP). If the semileptonic $B$ 
decay anomalies are indeed due to some NP, such operators can potentially lead to LFV decays involving the second 
and the third generation leptons, like $\tau\to
3\mu$. In this paper, we explore how far the nature of NP can be unraveled at the next generation 
$B$-factories like Belle-II, provided the decay $\tau\to 3\mu$ has been observed. We use four observables 
with which the differentiation among NP operators may be achieved to a high confidence level. Possible 
presence of multiple NP operators are also analysed with the Optimal Observable technique. While the 
analysis can be improved even further if the final state muon polarisations are measured, we present this work
as a motivational tool for the experimentalists, as well as a template for the analysis of similar processes.

\end{abstract}



\setcounter{footnote}{0}
\renewcommand*{\thefootnote}{\arabic{footnote}}

\section{Introduction}

Lepton flavour is an accidental symmetry of the Standard Model (SM), and there are many extensions of the SM,
like the seesaw models, supersymmetric SM, flavour-changing $Z'$ or scalars, leptoquarks, 
or left-right symmetric models, that can
naturally break this symmetry. Even within the ambit of SM, neutrino mixing provides a 
source of leptonic flavour violation (LFV), but the rates are too small to be observed in any near future
\cite{LFVrates}. Thus, observation of any LFV decay is a smoking gun signal of New Physics (NP).  
In general four types of LFV processes have been looked for: (i) leptonic decays ($\tau\to 3e$, $\tau\to 3\mu$, 
$\mu\to 3e$, $\tau\to 1e+2\mu$, $\tau\to 1\mu+2e$), (ii) radiative decays ($\tau\to e\gamma$, $\tau\to\mu\gamma$, 
$\mu\to e\gamma$), (iii) semileptonic decays ($\ell_1\to \ell_2M$, where $M$ is some meson), and (iv) 
conversion (like $\mu \to e$). They are not all 
independent, {\em e.g.}, a flavour-changing electromagnetic penguin can also give rise to leptonic LFV 
decays. Most of the decays, however, 
have very stringent limits \cite{LFVrates}, the branching ratios (BR) being typically of the order of $10^{-8}$ 
or even smaller. 

The interest in such LFV decays have recently been rekindled from the observation that some of the 
semileptonic $B$ decay modes show anomalous deviations from the SM expectations, which may
possibly be explained by lepton flavour non-universality (LFNU) 
as well as LFV. As an example, let us refer the reader 
to a recent attempt in Refs.\ \cite{CKMS,CKMS2}, where the authors have shown that 
both $R_K, R_{K^*}$ and $R(D), R(D^*)$ anomalies can be explained satisfactorily with only two new 
operators, if the weak and mass bases of the charged leptons $\{\mu,\tau\}$ are related by a field rotation
\cite{GGL}. The apparent excess in the LFV decay channel of the Higgs boson,
$h\to\mu\tau$, as was once reported by the CMS Collaboration \cite{1502.07400}, coupled with such $B$ decay 
anomalies, could lead to some well-motivated and fairly constrained models of LFV
\cite{Choudhury:2016ulr}. 
The LFV operators with only leptonic fields cam also be induced by Renormalisation Group (RG) running 
of semileptonic operators \cite{feruglio}. 
Thus, there is enough motivation to seriously look into such LFV channels; they 
might be observed at the Large Hadron Collider (LHC) \cite{lhc-mutau}, 
or dedicated super-B factories like Belle-II. Implications of such LFV decays may also be found in
Ref.\ \cite{request}. 

In this paper, we will focus solely on the leptonic channel $\tau\to 3\mu$. 
The channel has already been studied in detail in the literature; there are both model-independent 
\cite{0711.0792,1506.07786,1511.07434} as well as partially model-dependent \cite{0802.0049,1701.00870} 
studies. The Belle Collaboration has an upper bound on the BR \cite{1001.3221} 
\beq
{\rm BR}(\tau\to 3\mu) < 2.1\times 10^{-8}
\label{e:bound}
\eeq
at 90\% confidence level (CL). One reason for looking at this particular channel is the possibility 
of leptonic rotation in the $\{\mu,\tau\}$ sector as mentioned above, which invariably leads to such LFV 
channels out of lepton flavour conserving operators in the weak basis. Another reason, of course, is
the relative ease with which the final state muons can be detected in both hadronic and $e^+e^-$ colliders.   

Here we would like to push the studies on LFV in $\tau\to 3\mu$ further by asking and answering a few 
questions. Observation of even a single $\tau\to 3\mu$ event is a definite signal for NP.
Assuming that one observes, possibly at a super-B factory like Belle-II,
a few events for the LFV decay in question, will one be able to unearth the nature of the 
possible operators that can lead to such a decay? 
It has been shown \cite{0711.0792,0802.0049} that there can be six independent
LFV operators in the chiral 
basis that lead to $\tau\to 3\mu$. If the final state muon polarisations are not measured, 
all the operators are {\em a priori} equally probable, and obviously only the number of events will not tell 
us anything about the presence or absence of any of these operators; it can only yield some estimate on 
the respective Wilson coefficients (WC). 
So, are there observables which will help us to differentiate between these operators? 
We will show that this is indeed possible, without using higher-order
differential cross-sections like Ref.\ \cite{0711.0792} which may have very few number of events in each bin. 
At this point, we also note that some more operators can be generated through Fierz reordering, 
but obviously they are not independent of the first six, and therefore we will not consider them any further. 

The second question that we would like to ask is whether the existence of more than one such operators can be
disentangled from the data. Here the answer will be partially positive, unless, again, the muon polarisations are
measured. If one can have a sizeable number of events, and measure the muon polarisations too, one 
may have in principle further observables, but we would like to be conservative. Anyway, as we will show,
one does not expect more than 70 events or so at the most with an integrated luminosity ${\cal L}_{\rm int} =
50$ ab$^{-1}$ at Belle-II. 

We will use the method of Optimal Observables (OO), which has already been used in different areas of 
particle physics \cite{gunion,atwood}, and in particular, for flavour physics \cite{S3,zc}. 
This method displays the amount of significance level (``how many sigma" in standard parlance) 
by which one point in the allowed parameter 
space can be separated from another point. This is the only way to approach the question of 
model differentiation before the arrival of the data. Once one has the data, other methods, like the 
unbinned multivariate maximum likelihood, may be employed. 

A related question is  the number of events with which one can have a successful differentiation among models,
where a model is specified by its operator structure and WCs. As expected, if the number of events is too small, 
it will be harder to differentiate among various models, or in other words, the significance level will be lower. 
We will quantify this statement subsequently.

The paper is arranged as follows. In Section 2, we enlist all the possible NP operators that can give rise to
the $\tau\to 3\mu$ decay, and the observables that we deal with are discussed in Section 3. 
In Section 4, we show the differentiation among models with only one NP
operator. Section 5 deals with models with two such NP operators, and we discuss how well the presence 
of the second operator can be found out from various observables. Section 6 summarizes and concludes 
the paper.

\section{The New Physics Operators}

For this section, we will follow the notation and convention of Ref.\ \cite{0802.0049}. The most general LFV 
Lagrangian can be written as
\bea
{\cal L} &=& \frac{1}{\Lambda^2}
\bigg[\gsll (\bar\mu_L\mu_R)(\bar\mu_R\tau_L) + \gslr (\bar\mu_L\mu_R)(\bar\mu_L\tau_R) + 
\gsrl (\bar\mu_R\mu_L)(\bar\mu_R\tau_L) + \gsrr (\bar\mu_R\mu_L)(\bar\mu_L\tau_R)  \nonumber\\
&& \left. +  \gvll (\bar\mu_R\gamma^\alpha\mu_R)(\bar\mu_L\gamma_\alpha\tau_L) + 
\gvlr (\bar\mu_R\gamma^\alpha\mu_R)(\bar\mu_R\gamma_\alpha\tau_R) \right.\nonumber\\
&& \left. +  \gvrl (\bar\mu_L\gamma^\alpha\mu_L)(\bar\mu_L\gamma_\alpha\tau_L) + 
\gvrr (\bar\mu_L\gamma^\alpha\mu_L)(\bar\mu_R\gamma_\alpha\tau_R) \right.\nonumber\\
&&  +  \frac12 g^T_{LR} (\bar\mu_L\sigma^{\alpha\beta}\mu_R)(\bar\mu_L\sigma_{\alpha\beta}\tau_R) + 
\frac12 g^T_{RL} (\bar\mu_R\sigma^{\alpha\beta}\mu_L)(\bar\mu_R\sigma_{\alpha\beta}\tau_L) \bigg]\,,
\label{lfvlag}
\eea 
and we will denote the operator accompanying $g^X_{IJ}$ ($X=S,V,T$, and $I,J=L,R$) as $O^X_{IJ}$. 
$\Lambda$ is the cutoff scale, which we will set at 5 TeV for our analysis. 
We separate the operators into three major classes: ${\sf S}$ (operators of the form 
$O^S_{IJ}$), ${\sf V}$ (the $O^V_{IJ}$ operators) and ${\sf T}$ (the tensor 
operators $O^T_{IJ}$). Thus, the effective Lagrangian is of the form
\beq
{\cal L} = \frac{1}{\Lambda^2} \left[ \sum_{I,J=L,R} \left(g^S_{IJ} O^S_{IJ} + g^V_{IJ} O^V_{IJ}\right)
+ \sum_{I\not=J} g^T_{IJ} O^T_{IJ}\right]\,.
\label{e:lag-conc}
\eeq 
In the above mentioned basis, not all the ten operators are independent; Fierz transformation relates 
the two tensor operators with the rest, and the pairs $O^S_{LL}$-$O^V_{LL}$ and 
$O^S_{RR}$-$O^V_{RR}$ are also related \cite{0711.0792}. Thus, only four scalar and two vector 
operators are enough to span the operator basis. 
However, we keep all of them for the time being, as the mediator that has been integrated out may give rise 
to operators that are linear combinations of the six independent ones, like 
the tensor operators that can be generated from some hypothetical spin-2 mediators. 

Writing the BR in terms of the new WCs \cite{0802.0049}, one may easily show that the decay 
$\tau \to 3\mu$ has maximal sensitivity to {\sf V} or {\sf T} operators.
As an example, the present bound on ${\rm BR}(\tau\to 3\mu) < 2.1\times 10^{-8}$ translates to 
$|g_{RL}^S| \approx 1$, while $|g_{RL}^T|, |g_{RL}^V| \sim {\cal O}(0.1)$. 
Thus, for a given number of events, the reach for {\sf V} or {\sf T} WCs is better than that for the {\sf S} 
ones.  

The left-chiral fields being $SU(2)$ doublets, one can also get a neutrino-antineutrino pair out of the operators 
$O^S_{RR}$ and $O^V_{RR}$, which technically gives an extra contribution to the SM $\tau$ decay channel 
$\tau\to \mu\bar\nu_\mu \nu_\tau$. However, the couplings will turn out to be so constrained as not to
affect this channel in any significant amount. Similar LFV operators for $\mu\to 3e$ may affect the 
extraction of the Fermi coupling $G_F$ from the muon lifetime in a measurable way. 

One may try to look for $\tau\to\mu\gamma$ by contracting a pair of muons and taking the photon
with momentum $q^\mu$ out from the loop. For the scalar operators, the contribution vanishes in the $q^2=0$ 
limit, and for the vector operators, this amounts to charge and not the transition magnetic moment 
renormalisation. 

To begin with, we will consider the presence of one operator at a time. This generates six independent 
models spanning over the ${\sf S}$ and ${\sf V}$ classes. Next, we will consider the presence of two 
operators at a time, which include the well-motivated combinations like
\bea
O^S_{9L} = O^S_{LL} + O^S_{RL}\,,  && O^S_{10L} = O^S_{LL} - O^S_{RL}\,,  \nonumber\\
O^S_{9R} = O^S_{LR} + O^S_{RR}\,, &&  O^S_{10R} = O^S_{LR} - O^S_{RR}\,,
\eea
and similarly for the ${\sf V}$ class. Our goal will be to pinpoint whether or not these two coupling scenarios 
can be differentiated from those involving only one coupling at a time.

\section{Observables}

In this section, we will define the observables which we have used in our analysis to differentiate the effects of different NP operators. 
As an example, we consider the ${\sf S}$ class of operators, $O^S_{IJ}$, taken from Eq.\ (\ref{e:lag-conc}), and 
consider the decay $\tau^-\to \mu^-\mu^+\mu^-$.
The double differential cross-section for the antimuon is given, after integrating out the 
phase space for the two muons, by
\begin{align}
\frac{d B_\tau}{dx\, d(\cos\theta)} &= \frac{T_\tau\, m_\tau^5}{ 128 \times 48\pi^3 \Lambda^4}\left[ 
3x^2 (|g_{RL}^S|^2 +|g_{LL}^S|^2+|g_{RR}^S|^2+|g_{LR}^S|^2)\right.\nn\\
&\left.-x^3 (3|g_{RL}^S|^2+2|g_{LL}^S|^2+2|g_{RR}^S|^2 + 3 |g_{LR}^S|^2)\right. \nn \\
&\left.+\,x^2 \cos\theta\,(3 |g_{RL}^S|^2 + |g_{LL}^S|^2 - |g_{RR}^S|^2 - 3|g_{LR}^S|^2 )\right.\nn\\
&\left.- x^3 \cos\theta\,(3|g_{RL}^S|^2+ 2|g_{LL}^S|^2- 2|g_{RR}^S|^2- 3 |g_{LR}^S|^2)\right]\,,
\label{e:dgdx}
\end{align}
where we take the muons to be massless, and use the notation
\beq
B_\tau \equiv {\rm BR} (\tau\to 3\mu)\,.
\eeq
Here, $T_\tau = 1/\Gamma$ is the lifetime of the $\tau$ lepton,
$x = 2E_{\bar\mu}/m_\tau$ is the reduced energy of the
antimuon, and $\theta$ is angle between the polarization of the $\tau$ and the momentum of the antimuon,
following the convention of Ref.\ \cite{1506.07786}. For further discussion, let us define
\bea
g_1 &\equiv& |g_{RL}^S|^2 +|g_{LL}^S|^2+|g_{RR}^S|^2+|g_{LR}^S|^2\,,\nonumber\\ 
g_2 &\equiv& 3|g_{RL}^S|^2+2|g_{LL}^S|^2+2|g_{RR}^S|^2 + 3 |g_{LR}^S|^2\,,\nonumber\\
g_3 &\equiv& 3 |g_{RL}^S|^2 + |g_{LL}^S|^2 - |g_{RR}^S|^2 - 3|g_{LR}^S|^2\,, \nonumber\\
g_4 &\equiv& 3|g_{RL}^S|^2+ 2|g_{LL}^S|^2- 2|g_{RR}^S|^2- 3 |g_{LR}^S|^2\,.
\eea
Thus
\beq
B_\tau = \frac{T_\tau\, m_\tau^5}{ 128 \times 24\pi^3 \Lambda^4}\left[g_1 - \frac14 g_2\right]\,.
\label{brscalar}
\eeq
The number of events gives the information only on the combination $g_1-\frac14 g_2$. 

Another observable is the observed integrated forward-backward asymmetry $A_{FB}$, defined as 
\beq
A_{FB} = \frac{N_F - N_B}{N_F + N_B}\,,
\label{e:afb}
\eeq
with
\bea
N_F &=& \sigma_{\rm Prod} {\cal L}_{\rm int}\epsilon\, \int_{0}^{1}\, dx\, \int_{0}^{1}\, d(\cos\theta) 
\frac{dB_\tau}{dx\, d(\cos\theta)}\,,\nonumber\\
N_B &=& \sigma_{\rm Prod} {\cal L}_{\rm int} \epsilon\, \int_{0}^{1}\, dx\, \int_{-1}^{0}\, d(\cos\theta) 
\frac{dB_\tau}{dx\, d(\cos\theta)}\,,
\eea
where $\sigma_{\rm Prod}$ is the $\tau$ production cross-section, ${\cal L}_{\rm int}$ is the integrated luminosity, and 
$\epsilon$ is the combined detection efficiency in the $\tau\to 3\mu$ channel. 

We will also define the $x$-dependent asymmetry, normalised to the total decay width, as
\beq
\frac{dA_{FB}}{dx} \equiv A'_{FB}(x) = \sigma_{\rm Prod} {\cal L}_{\rm int} \epsilon\,
\frac{\int_{0}^{1}\, d(\cos\theta)\,  \frac{dB_\tau}{dx\, d(\cos\theta)} -
\int_{-1}^{0}\, d(\cos\theta)\, \frac{dB_\tau}{dx\, d(\cos\theta)}}{N_B + N_F}
\equiv \frac{N_F(x) - N_B(x)}{N}\,,
\label{afbx1}
\eeq
where $N=N_F + N_B$ gives the total number of signal events.

Instead of the antimuon, one can play an identical game with one of the same-sign muons
({\em i.e.}, the one with the same sign as the decaying $\tau$), say the more 
energetic of the two. Let us define
\bea
g'_1 &\equiv& |g_{RL}^S|^2 + 3 |g_{LL}^S|^2+ 3 |g_{RR}^S|^2+ |g_{LR}^S|^2\,,\nonumber\\ 
g'_2 &\equiv& |g_{RL}^S|^2+ 4 |g_{LL}^S|^2+ 4 |g_{RR}^S|^2 + |g_{LR}^S|^2\,,\nonumber\\
g'_3 &\equiv& |g_{RL}^S|^2 + 7 |g_{LL}^S|^2 - 7  |g_{RR}^S|^2 - |g_{LR}^S|^2\,, \nonumber\\
g'_4 &\equiv& |g_{RL}^S|^2+ 4 |g_{LL}^S|^2- 4 |g_{RR}^S|^2-  |g_{LR}^S|^2\,,
\eea
so that
\beq
\frac{dB_\tau}{dy\, d(\cos\alpha)} = 
\frac{T_\tau\, m_\tau^5}{128 \times 96\pi^3 \Lambda^4}\left[ 
3y^2 g'_1 - 2y^3 g'_2 + y^2 \cos\alpha\, g'_3 -2y^3\cos\alpha\, g'_4\right]\,,
\label{e:dgdy}
\eeq
where
$y = 2E_{\mu}/m_\tau$ is the reduced energy of the more energetic same-sign muon, 
and $\alpha$ is angle between its direction and the polarisation of the $\tau$ lepton.

In an analogous way to Eq.\ (\ref{afbx1}), one can define 
\beq
{\cal A}'_{FB}(y) = 
\frac{N_F(y)-N_B(y)}{N}=
\sigma_{\rm Prod} {\cal L}_{\rm int} \epsilon\,
\frac{\int_{0}^{1}\, d(\cos\alpha)\,  \frac{dB_\tau}{dy\, d(\cos\alpha)} -
\int_{-1}^{0}\, d(\cos\alpha)\, \frac{dB_\tau}{dy\, d(\cos\alpha)}}{N}.
\label{afby1}
\eeq
As will be shown later, the observables $A_{FB}^{\prime}$ and ${\cal A}_{FB}^{\prime}$
are useful to differentiate the sensitivities of the subtypes of operators within a particular type,
say {\sf S} or {\sf V}. 

\begin{figure}[htbp]
\centering
   \subfloat[\label{brvsgrlzoomed}]
  {\includegraphics[height=5.5cm]{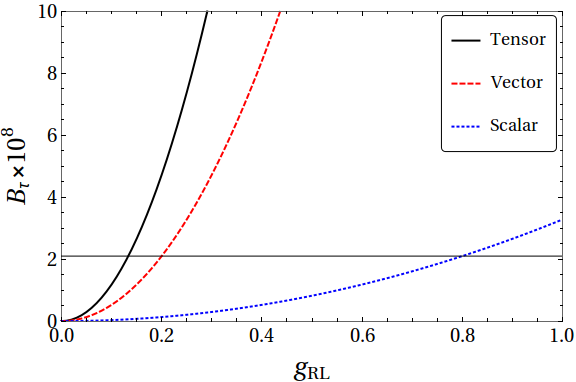}}
  \subfloat[\label{brvsgrrzoomed}]
  {\includegraphics[height=5.5cm]{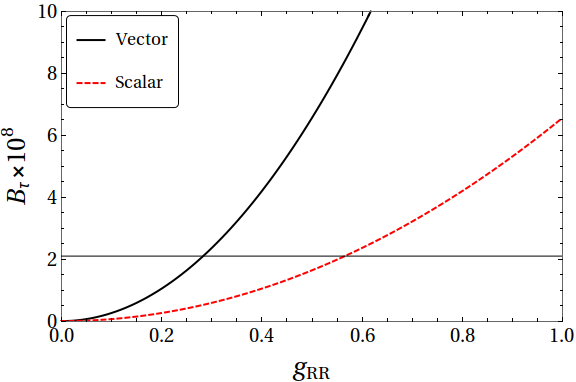}}
  \caption{ (a) Variation of ${\rm BR}(\tau \to 3\mu)$ with the WCs $g_{RL}^I$ ($I = S,V,T$).
  (b) The same for $g_{RR}^I$ ($I = S,V$). The horizontal line shows the present limit. The results for $g^I_{LR}$
  are identical to those for $g^I_{RL}$, and the results for $g^I_{LL}$ are identical to those for $g^I_{RR}$.}
\label{f:brvswc}
\end{figure}

Similarly, for the {\sf V} class of models, one obtains, in an analogous way to Eq.\ (\ref{e:dgdx}),   
\begin{align}
\frac{d B_\tau}{dx\, d(\cos\theta)} &= \frac{T_\tau\, m_\tau^5}{ 128 \times 12\pi^3 \Lambda^4}\left[ 
3x^2 (4|g_{RL}^V|^2 +|g_{LL}^V|^2+|g_{RR}^V|^2+4|g_{LR}^V|^2)\right.\nn\\
&\left.-2x^3 (6|g_{RL}^V|^2+|g_{LL}^V|^2+|g_{RR}^V|^2 + 6 |g_{LR}^V|^2)\right. \nn \\
&\left.+ x^2 \cos\theta(12 |g_{RL}^V|^2 + |g_{LL}^V|^2 - |g_{RR}^V|^2 - 12|g_{LR}^V|^2 )\right.\nn\\
&\left.- 2x^3 \cos\theta\,(6|g_{RL}^V|^2+ |g_{LL}^V|^2- |g_{RR}^V|^2- 6 |g_{LR}^V|^2)\right]\,.
\label{e:dgdxv}
\end{align}
The corresponding BR is
\beq
B_\tau = \frac{T_\tau\, m_\tau^5}{ 128 \times 12\pi^3 \Lambda^4}\left[2 g_1 -  g_2\right]\,,
\label{brvector}
\eeq
where,
\bea
g_1 &\equiv& 4|g_{RL}^V|^2 +|g_{LL}^V|^2+|g_{RR}^V|^2+4|g_{LR}^V|^2\,,\nonumber\\ 
g_2 &\equiv& 6|g_{RL}^V|^2+|g_{LL}^V|^2+|g_{RR}^V|^2 + 6 |g_{LR}^V|^2\,,\nonumber\\
g_3 &\equiv& 12 |g_{RL}^V|^2 + |g_{LL}^V|^2 - |g_{RR}^V|^2 - 12|g_{LR}^V|^2\,, \nonumber\\
g_4 &\equiv& 6|g_{RL}^V|^2+ |g_{LL}^V|^2- |g_{RR}^V|^2- 6 |g_{LR}^V|^2\,.
\eea
From Eqs.\ (\ref{e:dgdx}), (\ref{brscalar}), (\ref{e:dgdxv}), and (\ref{brvector}), one finds that {\sf V}-type 
operators generate more events than {\sf S}-type operators, if the orders of magnitude of their WCs are similar. 
The number of events as well as the angular distribution depend on the model subtype. We refer the 
reader to Fig.\ \ref{f:brvswc}, which shows this explicitly. 

For the same-sign muon, one gets
\beq
\frac{dB_\tau}{dy\, d(\cos\alpha)} = 
\frac{T_\tau\, m_\tau^5}{128 \times 24\pi^3 \Lambda^4}\left[ 
3y^2 g'_1 - 8y^3 g'_2 + y^2 \cos\alpha\, g'_3 - 8y^3\cos\alpha\, g'_4\right]\,,
\eeq
with 
\bea
g'_1 &\equiv& 4|g_{RL}^V|^2 + 3|g_{LL}^V|^2+ 3|g_{RR}^V|^2+ 4|g_{LR}^V|^2\,,\nonumber\\ 
g'_2 &\equiv& |g_{RL}^V|^2+ |g_{LL}^V|^2 + |g_{RR}^V|^2 + |g_{LR}^V|^2\,,\nonumber\\
g'_3 &\equiv& 4 |g_{RL}^V|^2 + 7|g_{LL}^V|^2 -7 |g_{RR}^V|^2 - 4|g_{LR}^V|^2\,, \nonumber\\
g'_4 &\equiv& |g_{RL}^V|^2+ |g_{LL}^V|^2- |g_{RR}^V|^2-  |g_{LR}^V|^2\,.
\eea
In an analogous way to Eqs.\ (\ref{afbx1}) and (\ref{afby1}), one can define $A'_{FB}(x)$ and 
${\cal A}'_{FB}(y)$.

While we will not discuss the {\sf T}-type operators separately, 
the double differential decay distribution is given by,
\beq
\frac{d B_\tau}{dx\, d(\cos\theta)} = \frac{T_\tau\, m_\tau^5}{ 128 \times 4\pi^3 \Lambda^4}
\, 9 x^2(1-x)\, \left[g_1
+ g_2 \cos\theta\right]\,,
\eeq
where,
\beq
g_1 \equiv |g_{RL}^T|^2+ |g_{LR}^T|^2\,,\ \ 
g_2 \equiv |g_{RL}^T|^2- |g_{LR}^T|^2\,.
\eeq
Thus
\beq
B_\tau = \frac{3 T_\tau\, m_\tau^5}{ 128 \times 8\pi^3 \Lambda^4}\, g_1\,.
\eeq

\section{Analysis}

In this section, we discuss the current and future sensitivities of the {\sf S}, {\sf V}, and {\sf T} 
operators on the observables $B_{\tau}$, $A'_{FB}$. and ${\cal A}'_{FB}$. The next subsection deals 
with the simplified cases where only one operator is considered to be present at a time. While this 
is instructive and sheds a lot of light on the differentiating power of the observables, a more realistic 
scenario might involve more than one operators. Thus, in the next Section, we discuss the cases 
where two operators are simultaneously present, and try to see whether those two-operator models can 
be separated from the single-operator ones. 

\subsection{One Operator Models}

Let us assume, to start with, that only one out of the ten possible operators shown in Eq.\ (\ref{lfvlag}) is 
present, notwithstanding the fact that not all of them are mutually independent, some being related to the 
others through Fierz rearrangement. 
In Fig.\ \ref{f:brvswc}, we show how the branching ratio $B_{\tau}$ depends on the WCs $g_{RL}^{S,V,T}$ and 
$g_{RR}^{S,V}$. Identical plots are obtained if one replaces $LR$ with $RL$, and $RR$ with $LL$. This 
$R\leftrightarrow L$ symmetry is true for all subsequent observables and their distributions, which reduces 
possible independent cases worth discussing by a factor of 2.

Given the combination $IJ$, if theory tells us the approximate magnitude of the WC $g^X_{IJ}$, even with 
the number of events as the sole observable, one can almost immediately differentiate $X=S$ case with $X=V$
or $T$. With higher statistics even a differentiation between {\sf V} and {\sf T} may be possible. The present
limit on $B_\tau$ is translated to
$\vert g^T_{RL}\vert \leq 0.13$, $\vert g_{RL}^{V}\vert \leq 0.20$, $\vert g^S_{RL}\vert \leq 0.80$, 
$\vert g_{RR}^{V}\vert \leq 0.28$, and $\vert g^S_{RR}\vert \leq 0.57$. However, as we do not have any 
{\em a priori} knowledge of the magnitudes of the WCs, we have to look for some other observables and 
use the number of events as a normalisation. In other words, we will assume that the total number of events
has been given to the community by the experimentalists and see how much extra information we can
extract. 

\begin{figure}[t]
\begin{center}
 \includegraphics[height=4.5cm]{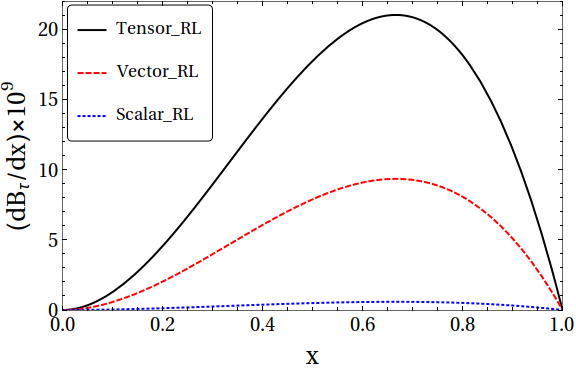}~~~
 \includegraphics[height=4.5cm]{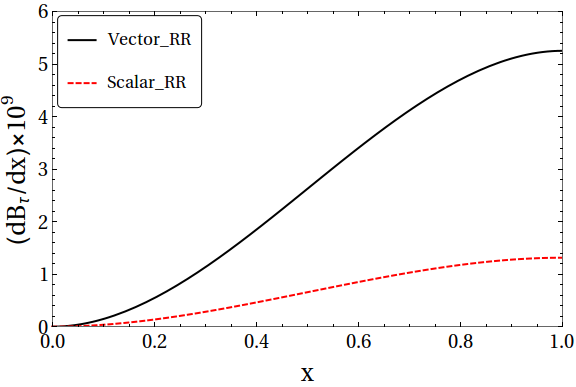}
\end{center}
\caption{The decay rate distribution $dB_\tau/dx$ for different operators, with the relevant
$g^{S,V,T}_{IJ} \approx 0.1$. }
\label{f:dbdx}
 \end{figure}

In Fig.\ \ref{f:dbdx}, we show how the differential rate $dB_\tau/dx$ varies with the muon energy variable $x$
for a fixed value of the WC, set at $0.1$.
With the normalisation included, the area under the curve gives the total number of events in different 
$x$-bins. Note that due to possible paucity of events, one may have only a few bins, 2 or 3, before the
data starts thinning out too much to have any statistical significance. Thus, the continuous distribution 
showed in Fig.\ \ref{f:dbdx} is an idealised scenario. Even then, what we find is that the number of events 
will be markedly different for different classes of operators if the WCs are of the same order, which is 
very much along the expected line. On the other hand, the asymmetry variables $A'_{FB}(x)$ or 
${\cal A}'_{FB}(y)$ must show identical pattern for all operators, {\sf S}, {\sf V},
or {\sf T}, with a fixed chirality structure, because the overall normalisation cancels in the ratio.   

The next task would be to differentiate among the various chiral subclasses of a particular class of model. 
For illustration, we will take the {\sf S} class of models, and consider the presence of one {\sf S}-type 
operators at one time. As the sensitivities are higher for {\sf V} and {\sf T} classes, whatever results one has
for {\sf S} will only be more enhanced and pronounced for other classes. At the same time, if the underlying
theory predicts $g_{IJ}$ values of the order of unity, very small number of events will be harder to sustain
under {\sf V} or {\sf T} classes.

\begin{figure}[t]
\centering
  \subfloat[\label{Ax_{fb}(binned) for 50 events}]
  {\includegraphics[height=4.5cm]{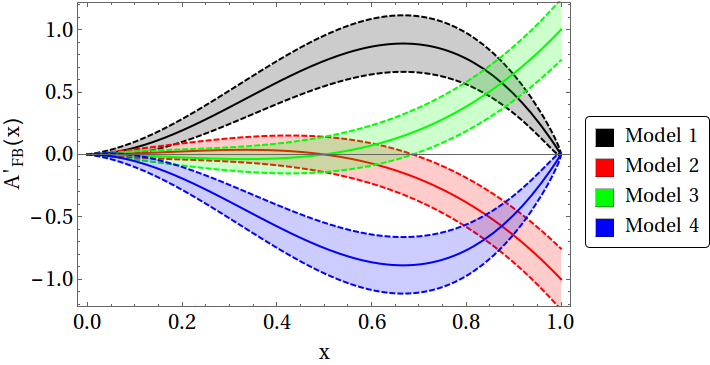}}
  \subfloat[\label{Ax_{fb}(binned) for 20 events}]
  {\includegraphics[height=4.5cm]{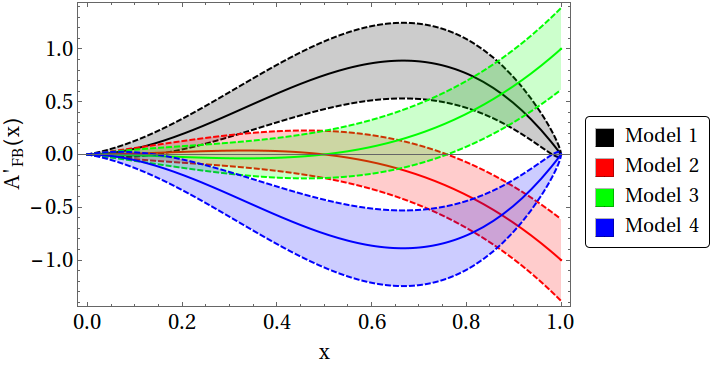}}
  
   \subfloat[\label{Ay_{fb}(binned) for 50 events}]
  {\includegraphics[height=4.5cm]{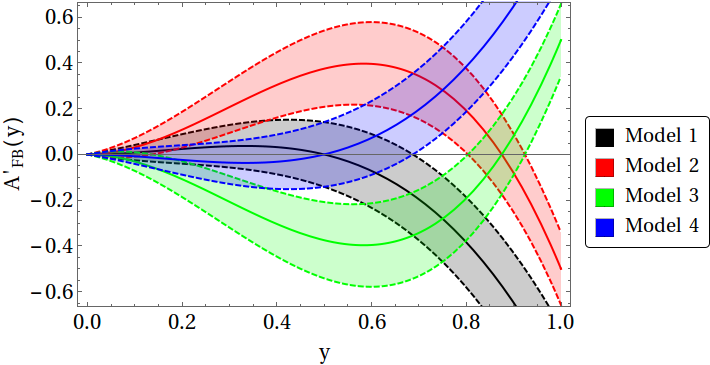}}
  \subfloat[\label{Ay_{fb}(binned) for 20 events}]
  {\includegraphics[height=4.5cm]{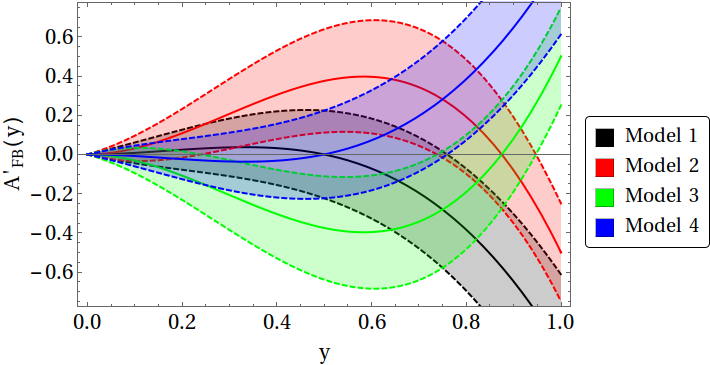}}
  \caption{ $A'_{FB}(x)$ for the antimuon with (a) 50 and (b) 20 events for the
  four single coupling {\sf S} class of models. The same for the more energetic of the two muons,
  ${\cal A}'_{FB}(y)$, with $($c$)$ 50 and (d) 20 events. 
}
\label{f:single}
\end{figure}

In the single-coupling scheme, we consider four different models, depending upon which operator contributes, and denote them as 
\beq
{\rm Model~1}: O^S_{RL}\,,\ \ {\rm Model~2}: O^S_{LL}\,,\ \ 
{\rm Model~3}: O^S_{RR}\,,\ \ {\rm Model~4}: O^S_{LR}\,.
\eeq 
If only one operator contributes, $A'_{FB}(x)$ becomes a function of $x$ only and does not depend 
on the magnitude of the WCs:
\beq
A'_{FB}(x)_{1} = -A'_{FB}(x)_{4} = 6(x^2-x^3)\,,\ \ 
A'_{FB}(x)_{2} = -A'_{FB}(x)_{3} = x^2-2x^3\,.
\eeq 
The integrated asymmetry $A_{FB}(i)$ for the $i$-th model can be obtained by integrating $x\in [0:1]$, and the values are 
\beq
A_{FB}(1) = -A_{FB}(4) = \frac12\,,\ \ 
A_{FB}(2) = -A_{FB}(3) = -\frac16\,.
\eeq
There is a zero crossing only for models 2 and 3 at $x=\frac12$.

Similarly, for the forward-backward asymmetry ${\cal A}'_{FB}(y)$, we find
\beq
{\cal A}'_{FB}(y)_{1} = -{\cal A}'_{FB}(y)_{4} = y^2 - 2y^3\,,\ \ 
{\cal A}'_{FB}(y)_{2} = -{\cal A}'_{FB}(y)_{3} = \frac12 (7y^2-8y^3)\,.
\eeq 
While all of them show zero-crossing, for the last two models such crossing occurs almost at the end of
the kinematic range at $y=7/8$. The integrated asymmetries are (for $y\in [0:1]$)
\beq
{\cal A}_{FB}(1) = {\cal A}_{FB}(3) = - {\cal A}_{FB}(2) = -{\cal A}_{FB}(4) = -\frac16\,.
\eeq

The $A'_{FB}(x)$s for different models are shown in Figs.\ 
\ref{Ax_{fb}(binned) for 50 events} and \ref{Ax_{fb}(binned) for 20 events}. Similarly, ${\cal A}'_{FB}(y)$s 
are shown in Figs.\ \ref{Ay_{fb}(binned) for 50 events} and \ref{Ay_{fb}(binned) for 20 events}. 
In these figures, every theoretical line has 
broadened out to a thick band, often overlapping with each other. This happens because the number of the events is 
limited. For every $x$ ($y$), the error margin in $A'_{FB}(x)$ is approximately given by
\beq
\delta A'_{FB} = \sqrt{\left(\frac{\partial A'_{FB}(x)}{\partial N_F(x) }\right)^2 (\delta N_F(x))^2 + 
\left(\frac{\partial A'_{FB}(x)}{\partial N_B(x) }\right)^2 (\delta N_B(x))^2
+ \left(\frac{\partial A'_{FB}(x)}{\partial N}\right)^2 (\delta N)^2}
\eeq
where
\beq
\delta N_F(x) = \sqrt{N_F(x)},\ \  \delta N_B(x) = \sqrt{N_B(x)}\,,
\eeq
are the statistical errors in the number of events in the forward and backward directions respectively,
and $\delta N = \sqrt{N}$.

The expression for ${\cal A}'_{FB}(y)$ is analogous. 
We have not considered the correlation between $N_F(x)$ and $N_B(x)$; depending on the sign 
of the correlation, the expression can be an overestimation or underestimation, 
but as we do not have any {\em a priori} 
knowledge of the distribution, it is better to stick to zero correlation. 
The bands in Fig.\ \ref{f:single} indicate the $1\sigma$ error margins. 
Clearly, the resolving power is much less for 20 events than with 50 events. 

Because of the probable paucity of events, the asymmetries may be measured
 only with a limited number of bins. But even with two bins, low-$x$ ($0<x<0.5$) and high-$x$ ($0.5<x<1$), 
 one should be able to differentiate between competing models.

The existing bound on $\tau\to 3\mu$ comes from the analysis of 782 fb$^{-1}$ data from the Belle
collaboration \cite{1001.3221}, and 468 fb$^{-1}$ data from the BaBar collaboration \cite{1002.4550}. 
With a production cross-section of $0.919$ nb for the $\tau^+\tau^-$ pairs, one gets 720 million such pairs
at Belle and 420 million at BaBar. For 50 ab$^{-1}$ of integrated luminosity at Belle-II, one expects
$N_P=4.6\times 10^{10}$ $\tau^+\tau^-$ pairs. With a detection efficiency of $7.6\%$ \cite{1001.3221},
and using the present bound given in Eq.\ (\ref{e:bound}), the maximum number of such events is about
73. For our discussion, we will use two scenarios: one with $N=20$ and the other with $N=50$.
Note that the errors are only statistical in nature. There may be other uncertainties, like fixing the direction 
of the $\tau$ polarisation, which will widen up the bands, but that effect is expected to be small with the 
$\tau$ detection ability of Belle-II.

\begin{figure}[t]
\centering
  \subfloat[\label{dbrdx for 50 events}]
  {\includegraphics[height=4.5cm]{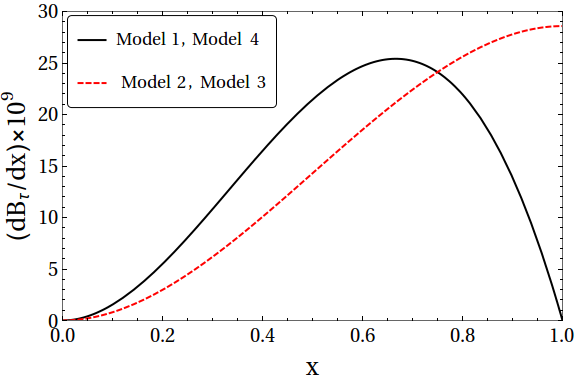}}
  \subfloat[\label{dbrdy for 50 events}]
  {\includegraphics[height=4.5cm]{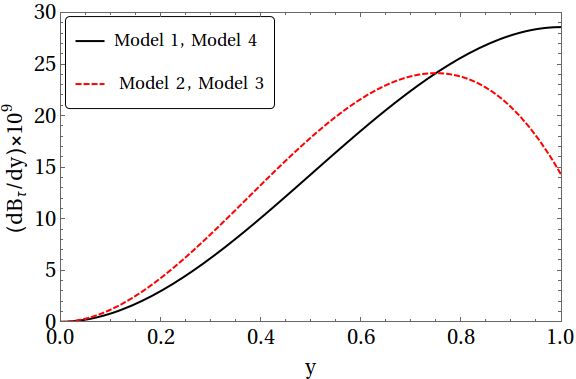}}

  \caption{ (a) $dB_\tau/dx$ and (b) $dB_\tau/dy$ for the four 
  single coupling {\sf S} class of models. 
}
\label{f:single2}
\end{figure}
 
If $A_{FB}$ turns out to be positive (negative), the viable models are 1 or 3 (2 or 4). Similarly, the 
positive (negative)-${\cal A}_{FB}$ models are 2 and 4 (1 and 3). Thus, measurement of only the sign of 
these asymmetries leave us with a twofold ambiguity. However, $x(y)$-dependent asymmetry measurements 
have the ability to resolve the same; {\sf S} and {\sf V}-type models behave identically. 
It is enough to measure the asymmetry in two bins: low $x(y)$ bin for $0\leq x(y) < \frac12$ and 
high $x(y)$ bin for $\frac12\leq x(y) \leq 1$. If these measurements were precise enough, it would have 
been sufficient not only to pinpoint the model but also to explore whether more than one operators are 
contributing \footnote{We, however, show a continuous $x(y)$-distribution of the asymmetries; one just needs 
to integrate over the respective bin to have an idea of the relative magnitudes.}. 
Unfortunately, with a limited number of events, the measurements cannot be that precise. 

Our results $dB_{\tau}/dx$ and $dB_{\tau}/dy$ in the single-coupling schemes are shown in 
Fig.\ \ref{f:single2}. The lines broaden out into bands if we take the errors and uncertainties into account.
Such broadening, in all probability, will make the lines indistinguishable from one another.  
However, all these models can be separated from each other 
from the asymmetry measurements, particularly in the high-$x(y)$ bin.

\section{Two Operator Models}

Once we establish that given enough events ($\sim 50$), it will be straightforward to differentiate between 
several one-operator models, the next question is: what if the data is not compatible with any of them? Note that
for the single-operator scheme to stand good, all the observables, and not only a few of them,
 have to be in the right ballpark specified by that model. 

However, even the principle of Occam's Razor may not be enough reason for not invoking the
double-operator scheme. 
We will, as before, be confined within the {\sf S} class of models,
and consider the cases where two WCs are nonzero at a time. As we have shown, this will be the case 
if the underlying theory forces the muon current to be pure $S$ or $P$ \footnote{And pure $V$ or $A$ 
for the {\sf V} class of models.}.  
Thus, the question we ask is: If the new physics is described by two 
$\tau\to 3\mu$ operators, with what confidence level can we differentiate that from those cases where only 
one of them is present? Note that the number of events will act as the tightest constraint on the parameter space. 
We will try to differentiate these models, hereafter called $O2$ for two effective operators, 
from that with only one operator, which we call the `seed' model. 
For example, we consider the following $O2$ models: 
\bea\label{modlst1}
{\rm Model~A} &:& O_{RL}^S~{\rm and}~O_{LL}^S\,,\nonumber\\
{\rm Model~B} &:& O_{RL}^S~{\rm and}~O_{RR}^S\,,\nonumber\\
{\rm Model~C} &:& O_{RL}^S~{\rm and}~O_{LR}^S\, .
\eea
We will compare the differentiability of these models (A, B and C) with the seed model with one operator
 $O^S_{RL}$. To achieve this goal, we need to find the parameter space spanned by $g^S_{RL}$ and 
another $g^S_{IJ}$, which depends on what $O2$ model we consider.
We further check with what confidence level 
the allowed regions for models A, B and C can be separated from the single-operator model given by 
$O^S_{RL}$. To complete the study, we include three further $O2$ models, namely,
\bea\label{modlst2}
{\rm Model~D} &:& O_{RR}^S~{\rm and}~O_{RL}^S\,,\nonumber\\
{\rm Model~E} &:& O_{RR}^S~{\rm and}~O_{LL}^S\,,\nonumber\\
{\rm Model~F} &:& O_{RR}^S~{\rm and}~O_{LR}^S\,,
\eea
where the first operator is treated as the seed. 

Models B and D are different, because of different seeds. The seed models are chosen in such 
a way as to have positive $A_{FB}$ for the opposite-sign muon for them; 
the negative $A_{FB}$ models will have a corresponding
relationship, which can be obtained by flipping $L$ and $R$, $L \leftrightarrow R$ \footnote{ 
If we consider the asymmetry based on the more energetic like-sign muon, models A-F have negative
asymmetry, while the corresponding $L\leftrightarrow R$ models have positive asymmetry.}.
Let us mention here that the confidence interval contours will depend on the type of seed operators being considered. 

For this part of the analysis, we will use the Optimal Observable (OO) technique. For a detailed discussion 
on this technique, we refer the reader to Refs.\ \cite{gunion,atwood}. In the context of $B$ decays, this 
method has been applied in Refs.\ \cite{S3,zc}. The essential point of the OO technique is that this gives the 
optimal set of observables (which are in general functions of experimental observables) with which two 
points in the parameter space of different models can be differentiated with maximum efficiency. In other
words, this gives the maximum possible separation, in terms of confidence level, between two points 
in the parameter space as a function of experimental observables. In practice, the systematic errors 
reduce the confidence intervals.

As has been shown in Refs.\ \cite{S3,zc}, this method is all the more useful when one does not have 
any experimental data; in the presence of data, one can do a maximum likelihood analysis. This also means
that not all the systematic uncertainties are taken into account. Thus, OO acts more as a motivational tool 
to the experimentalists than as an instrument for detailed quantitative theoretical studies. 

Even with only the {\sf S} class of operators, the parameter space of WCs is four-dimensional. A complete 
analysis is not only cumbersome but also of very little help in the real-life scenario where the number of 
events will definitely be below 100 and therefore a fine scan of the parameter space, with a two-dimensional
binning on $x$($y$) and $\cos\theta(\cos\alpha)$, will have so few events per bin as to make the analysis
meaningless. The only constraint on the WCs comes from the non-observability of the decay.



In the OO technique, one writes any observable ${\cal O}$, depending on a variable $\phi$, as
\beq
{\cal O}(\phi) = \sum_i C_i f_i(\phi)\,,
\label{e:def-oo}
\eeq
which can be generalised to a set of variables denoted by $\bm{\phi}$. Here, all the $f_i$s are independent, and $C_i$s are some constants. 
The major goal of this technique is to extract $C_i$s. In our case, $C_i$s will be functions of $g_i$s and $g'_i$s defined earlier. 
Our analysis can be done by defining a quantity analogous to $\chi^2$, such as 
\begin{align}\label{chidef}
 \chi^2 &= \sum_{i,j} (C_i - C_i^0)  V_{ij}^{-1}(C_j - C_j^0) .
\end{align}
The $C_i^0$s are called the seed values, which can be considered as model inputs.
The covariance matrix $V_{ij}$s are defined as
\begin{equation}
V_{ij} = \bra \Delta C_i \Delta C_j\ket = \frac{ (M^{-1})_{ij} \sigma_T}{N}\,, \text{with}\ \  
M_{ij} = \int \frac{f_i(\phi)f_j(\phi) }{O(\phi)} \, d\phi.
\label{vijdef}
\end{equation}
In the above expression of $V_{ij}$, $\sigma_T = \int O(\phi)\, d\phi$ and $N = \sigma_{\rm Prod}{\cal L}_{\rm int}
\epsilon B_\tau$, as defined earlier. For a specific model, $\chi^2$ gives the
confidence level separation between the seed value $C_i^0$ (seed model) and the model under consideration, 
parametrized by $C_i$.

Looking at Eq. (\ref{chidef}), it is clear that the shape of the fixed $\chi^2$ hypersurface depends on $V^{-1}_{i j}$, 
and the centroid of that (where $\chi^2 =\chi^2_{min} = 0$) changes with the seed values. These fixed $\chi^2$
surfaces are what determines the separation between models essentially. 
Thus separation between any two models 1 and 2, with seed at 1, will in general be not equal to the separation 
when the seed is at 2. This is the reason for treating Models B and D separately in 
Eqs.\ (\ref{modlst1}) and (\ref{modlst2}).

In the case of single operator model, the seed values of the WCs corresponding to the 50 events are obtained as  
\beq
|g^S_{RL}|^2 = 0.44~{\rm (A,B,C)}\,,\ \ \ 
|g^S_{RR}|^2 = 0.22~{\rm (D,E,F)}\,.
\eeq
For the negative $A_{FB}$ models, one may take $|g^S_{LR}|^2=0.44$ and $|g^S_{LL}|^2 = 
0.22$. Our additional inputs are 
\beq
m_\tau = 1.78~{\rm GeV}\,,\ \ T_\tau= 290.3~{\rm fs}\,, \ \ \Lambda = 5~{\rm TeV}\,.
\eeq

We show our results for the ${\sf S}$ class of models; ${\sf V}$ class of models will show identical 
results. The observables that we use are $A'_{FB}(x)$, ${\cal A}'_{FB}(y)$ (both defined in the previous section), as well as
$dB_\tau/dx$ and $dB_\tau/dy$, the expressions for which can be obtained from Eqs.\ (\ref{e:dgdx}) and (\ref{e:dgdy}):
\bea
\frac{dB_\tau}{dx} &=& 
\frac{m_\tau^5\, T_\tau}{ 128 \times 24\pi^3 \Lambda^4}(3x^2 g_1 - x^3 g_2)\,,\nonumber\\
\frac{dB_\tau}{dy} &=& 
\frac{m_\tau^5 T_\tau}{ 128 \times 48\pi^3 \Lambda^4}(3y^2 g'_1 - 2y^3 g'_2)\,.
\eea
From Eq.\ (\ref{afbx1}), one can write
\beq
A'_{FB}(x) = \frac{1}{B_\tau}\times T_\tau \frac{m_\tau^5}{ 128 \times 48\pi^3 \Lambda^4}(x^2 g_3 - x^3 g_4)\,.
\eeq
For 50 events, $B_\tau$ = $1.43\times 10^{-8}$. 
Similarly, 
\beq
{\cal A}'_{FB}(y) = \frac{1}{B_\tau}\times T_\tau \frac{m_\tau^5}{ 128 \times 96\pi^3 \Lambda^4}
(y^2 g'_3 - 2y^3 g'_4)\,.
\label{afby}
\eeq

Before we show our results for all the 6 models, let us mention a few important points here. 
\begin{enumerate}

 \item The determination of $\chi^2$ involves an integration over the variable $\phi$ of Eq.\ 
 (\ref{e:def-oo}). If over the region of integration the observable for the seed model becomes zero 
 for any value of $\phi$, the integration diverges. Thus, one has to cut off such badly behaving regions. 
 For example, if the observable for the seed model becomes zero at the end points, say $a$ and $b$, 
 one has to perform the integral between $a+\epsilon$ and $b-\epsilon$, where $\epsilon$ is taken to be so small 
 as not to affect the observable (like, say, the number of events). More concrete examples are given 
 below.
 
 \item One may ask why we do not use a two-variable analysis and use the double differential cross-section
 as the observable. This would have certainly been useful, and more powerful as an analytical tool, if we could
 manage a large number of events so that even the two-dimensional bins have enough number of events. 
 With a small number of events, such an analysis would not give much useful information.
 \end{enumerate}
 
 All our observables depend on only two functions, $f_1$ and $f_2$, with the argument being $x$ for 
 the opposite-sign muon and $y$ for the like-sign more energetic muon. 
 Depending on the observables, the combinations $C_1$ and $C_2$ are as follows.
 
\subsection{Observable: $A'_{FB}(x)$} 

\begin{figure}[t]
\centering
  \subfloat[\label{RL-LL}]
  {\includegraphics[height=5cm]{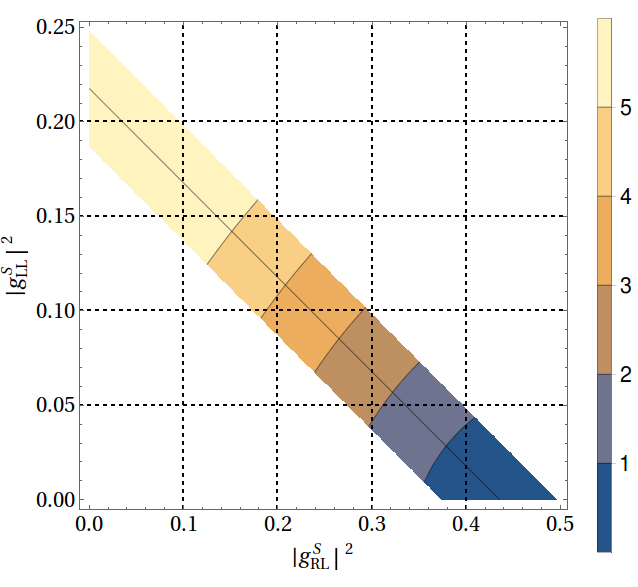}}
  \subfloat[\label{RL-RR}]
  {\includegraphics[height=5cm]{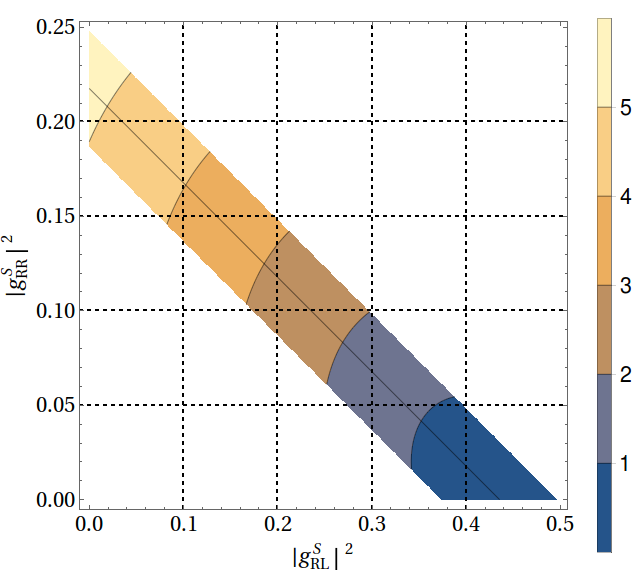}}
  \subfloat[\label{RL-LR}]
  {\includegraphics[height=5cm]{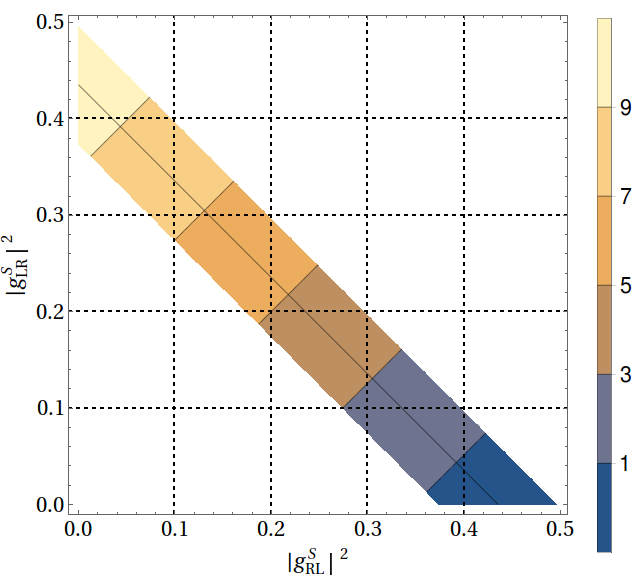}}  \\
  \subfloat[\label{RR-RL}]
  {\includegraphics[height=5cm]{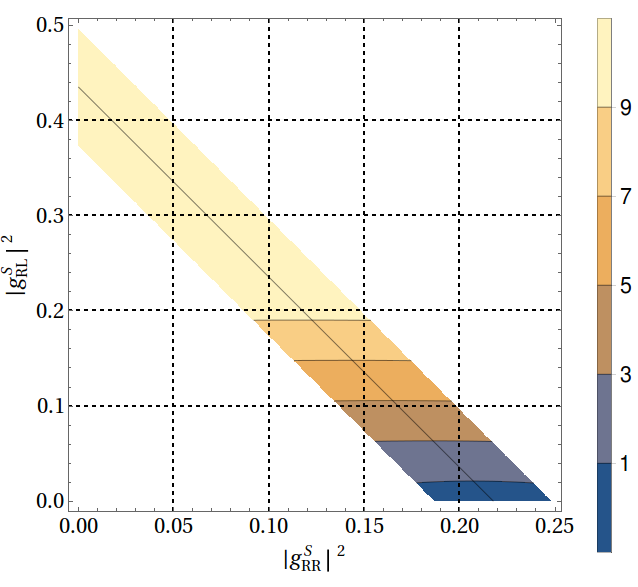}}
  \subfloat[\label{RR-LL}]
  {\includegraphics[height=5cm]{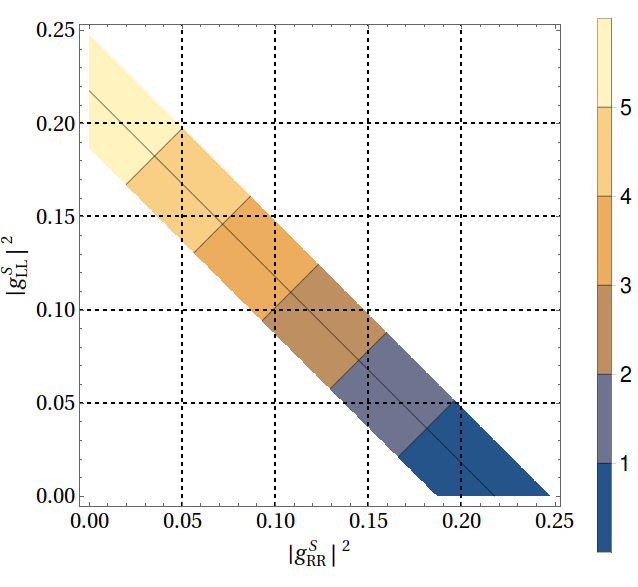}}
  \subfloat[\label{RR-LR}]
  {\includegraphics[height=5cm]{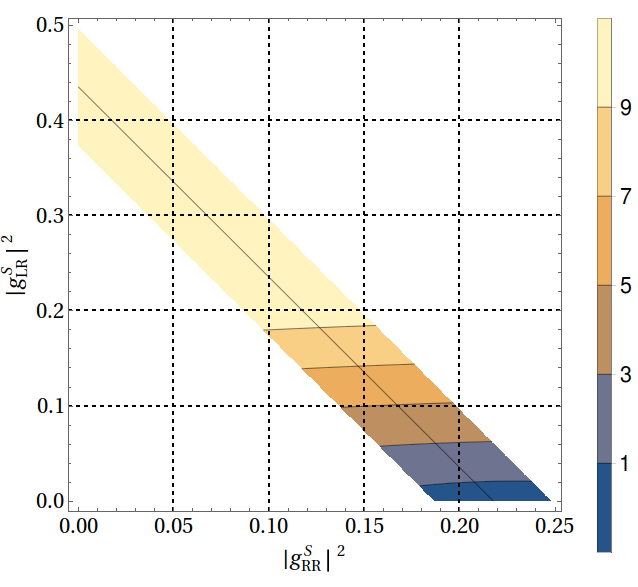}}
  \caption{The differentiability of the models A-F, shown in (a)-(f) respectively, from the 
  `seed' model, with $A'_{FB}(x)$ as the observable.}
\label{f:afbx}
\end{figure}

\begin{table}[htbp]
 \begin{center}
 \begin{tabular}{|c|c|c|c|c|}
 \hline
 Model & Seed & Second operator & $C_1$ & $C_2$\\
 \hline
 A & $RL$ & $LL$ & $3 |g_{RL}^S|^2 + |g_{LL}^S|^2$ & $3 |g_{RL}^S|^2 + 2 |g_{LL}^S|^2$\\
 B & $RL$ & $RR$ & $3 |g_{RL}^S|^2 - |g_{RR}^S|^2$ & $3 |g_{RL}^S|^2 - 2 |g_{RR}^S|^2$\\
 C & $RL$ & $LR$ & $3 |g_{RL}^S|^2 - 3 |g_{LR}^S|^2$ & $3 |g_{RL}^S|^2 - 3 |g_{LR}^S|^2$\\
 D & $RR$ & $RL$ & $ - |g_{RR}^S|^2 + 3 |g_{RL}^S|^2$ & $-2 |g_{RR}^S|^2 + 3 |g_{RL}^S|^2$\\
 E & $RR$ & $LL$ & $ - |g_{RR}^S|^2 + |g_{LL}^S|^2$ & $-2 |g_{RR}^S|^2 + 2 |g_{LL}^S|^2$\\
 F & $RR$ & $LR$ & $ - |g_{RR}^S|^2 - 3 |g_{LR}^S|^2$ & $-2 |g_{RR}^S|^2 - 3 |g_{LR}^S|^2$\\
 \hline
  \end{tabular}
  \caption{$C_1$ and $C_2$ for different models. The observable is $A'_{FB}(x)$.}
 \label{tab-afbx}
 \end{center}
 \end{table}

We show the coefficients $C_1$ and $C_2$ in Table \ref{tab-afbx}, and
Fig.\ \ref{f:afbx} displays our results for Models A-F. As the plots are not self-explanatory, 
let us clearly specify what they mean. The diagonal band with negative slope, in each of the plots, 
represents the allowed region in the parameter space of the various $O2$ models. Only the two relevant 
WCs are taken to be nonzero, keeping the others fixed at zero. 
Once the experimentalists obtain a certain number of events, this will specify a line in the 
two-dimensional parameter space over which the allowed models, each of them specified by some WCs, 
may lie. The exact position of the line will depend on what model one chooses, 
but the analysis must take into account the constraint imposed by this line \footnote{That is why
even for a two-parameter model the degree of freedom is only 1.}. The uncertainties in the data
will broaden the line to a band, whose width will ultimately depend on the number of events as well as the 
detector parameters. As a very rough guess, we take $\sqrt{N}$ to be the width for the line with $N$
events. The plots are drawn for $N=50$; thus, the band includes all the points for which the number of events 
lie approximately between 43 and 57. The separation contours are drawn on these bands only. We expect the 
bands to be narrower in actual experiments. 

Let us consider Fig.\ \ref{RL-LL}. This takes $|g^S_{RL}|^2 = 0.435$ as the seed value. The plot 
tells us that this one coupling model can be differentiated from the one with $|g^S_{RL}|^2 = 0.2$ and $|g^S_{LL}|^2 = 0.1$  
at more than $3\sigma$ if we have approximately 50 events and use $A'_{FB}(x)$ as our observable. 
Similarly, the model with $|g^S_{RL}|^2 = 0.35$ and 
$|g^S_{LL}|^2 = 0.05$ can be separated from the above mentioned seed model by less than $2\sigma$.
The actual numbers should be even worse as the 
systematic uncertainties will also creep in. Similar conclusions hold for Models B, C, D, E and F, for which 
the results are shown from Figs. \ref{RL-RR} to \ref{RR-LR} respectively. 
As we mentioned before, contours for Models B and D are not the same, although they involve the same 
set of operators. This is because the seed is different, which ultimately control the correlation matrix. 

For models D-F, the seed model has a zero crossing for $A'_{FB}(x)$ at $x=\frac12$. Unlike in the case 
of the differential decay distribution and observables proportional to it, the integrated observable in this case 
may become negative in different parts of the parameter space. This makes the covariance matrix $V_{i j}$ 
not positive definite. We note here that for our purpose, {\em i.e.} to construct $\chi^2$, the integrated 
observables serve only as the normalization of $V^{-1}_{i j} (= M_{i j})$. We have taken the modulus of the 
integrand for each value of $x (y)$ for this reason. 
This, while keeping the nature of the error ellipsoids intact, will always keep the covariance matrix 
positive definite. On the flip-side, this makes the integral diverge at the zero crossing point. Thus, to evaluate 
the correlation matrix $V_{ij}^{-1}$ and to cancel this divergence, one has to remove the tiny 
patch $0.495 < x < 0.505$ from integration. This 
has only a negligible effect on the number of events, but keeps the necessary integrations convergent. 

\begin{figure}[t]
\centering
  \subfloat[\label{2coupdist1}]
  {\includegraphics[height=5cm]{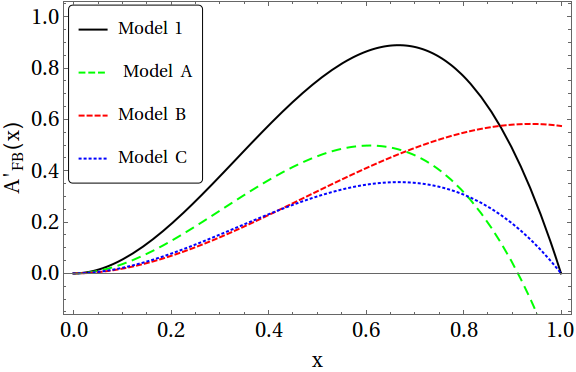}}
  \subfloat[\label{2coupdist2}]
  {\includegraphics[height=5cm]{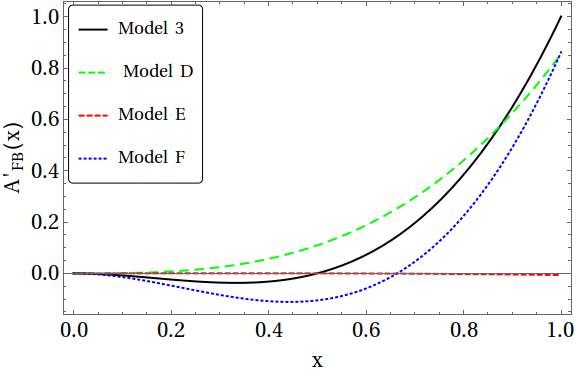}}
  \caption{Comparison of $A'_{FB}(x)$ for the $O2$ Models A-F with the 
  respective `seed' model. The relevant WCs are taken from Fig.\ \ref{f:afbx}.}
\label{2cdafbx}
\end{figure}

\begin{table}[htbp]
 \begin{center}
 \begin{tabular}{|c|c|c|c|c|c|}
  \hline
 Model  &  $|g_{RL}^S|^2$  & $|g_{LL}^S|^2$  & $|g_{RR}^S|^2$ & $|g_{LR}^S|^2$ & $A_{FB}$ (int.) \\
  \hline
1(Seed) &      0.435       &      -          & 	   -          &     -          & 0.5     \\
 \hline
      A &      0.265       &     0.085       & 	   -          &     -          & 0.239    \\
 \hline
      B &      0.186       &        -        & 	   0.125      &     -          & 0.309    \\
 \hline
      C &      0.305       &        -        & 	    -         &   0.131        & 0.2    \\
 \hline
3(Seed) &        -         &        -        & 	   0.218      &     -          & 0.167    \\
 \hline
      D &      0.063       &        -        & 	   0.186      &     -          & 0.215    \\
 \hline
      E &        -         &      0.110     & 	   0.108      &     -          & -0.001    \\
 \hline
      F &        -         &         -       & 	   0.188     &     0.061      & 0.073    \\
 \hline
  \end{tabular}
 \end{center}
 \caption{The WCs, as obtained from Fig.\ \ref{f:afbx}, 
 for which Models A-F are separable from the 
 respective seed models at $3\sigma$ level. The integrated asymmetries are also shown.}
 \label{t:2cdafbx}
 \end{table}
 
In Fig. \ref{2cdafbx}, we show, as an illustration, the behaviour of $A'_{FB}(x)$ for Models A-C 
vis-a-vis the seed Model 1 and Models D-F with seed Model 2, for which the differentiability is at the 
$3\sigma$ level. The corresponding WCs, extracted from Fig.\ \ref{f:afbx}, are displayed in Table \ref{t:2cdafbx}. 
We note that Models A-C can be differentiated from the seed Model 1 with only $|g^S_{RL}|^2$ 
for all values of $x$. On the other hand, Models D and F can be differentiated from seed Model 2
with only $|g^S_{RR}|^2$ (Model 3) only for medium values of $x$, and zero-crossing of $A'_{FB}(x)$ plays a crucial role.

\subsection{Observable: ${\cal A}'_{FB}(y)$ }

In an analogous way, one can use the more energetic of the like-sign muons, and the corresponding 
asymmetry ${\cal A}'_{FB}(y)$. The coefficients $C_1$ and $C_2$, from Eqs.\ (\ref{e:dgdy}) 
and (\ref{afby1}), are shown in Table \ref{tab-afby}. 

\begin{table}[htbp]
\begin{center}
 \begin{tabular}{|c|c|c|c|c|}
 \hline
 Model & Seed & Second operator & $C_1$ & $C_2$\\
\hline
 A & $RL$ & $LL$ & $ |g_{RL}^S|^2 + 7 |g_{LL}^S|^2$ & $ |g_{RL}^S|^2 + 4  |g_{LL}^S|^2$\\
 B & $RL$ & $RR$ & $|g_{RL}^S|^2 - 7 |g_{RR}^S|^2$ & $|g_{RL}^S|^2 - 4 |g_{RR}^S|^2$\\
 C & $RL$ & $LR$ & $|g_{RL}^S|^2 - |g_{LR}^S|^2$ & $|g_{RL}^S|^2 - |g_{LR}^S|^2$\\
 D & $RR$ & $RL$ & $ -7 |g_{RR}^S|^2 + |g_{RL}^S|^2$ & $ -4 |g_{RR}^S|^2 +  |g_{RL}^S|^2$\\
 E & $RR$ & $LL$ & $ -7 |g_{RR}^S|^2 + 7 |g_{LL}^S|^2$ & $ -4 |g_{RR}^S|^2 + 4 |g_{LL}^S|^2$\\
 F & $RR$ & $LR$ & $ - 7 |g_{RR}^S|^2 - |g_{LR}^S|^2$ & $ - 4 |g_{RR}^S|^2 -  |g_{LR}^S|^2$\\
 \hline
 \end{tabular}   
 \caption{$C_1$ and $C_2$ for different models, with ${\cal A}'_{FB}(y)$ as the observable.}
 \label{tab-afby}
 \end{center}
 \end{table}

\begin{table}[htbp]
 \begin{center}
 \begin{tabular}{|c|c|c|c|c|c|}
  \hline
 Model  &  $|g_{RL}^S|^2$  & $|g_{LL}^S|^2$  & $|g_{RR}^S|^2$ & $|g_{LR}^S|^2$ & $A_{FB} (int.)$ \\
  \hline
1(Seed) &      0.435       &      -          & 	   -          &     -          & -0.167     \\
 \hline
      A &      0.314       &     0.061       & 	   -          &     -          & -0.074    \\
 \hline
      B &      0.313       &        -        & 	   0.061      &     -          & -0.167    \\
 \hline
      C &      0.216       &        -        & 	    -         &   0.219        &  0.001    \\
 \hline
 3(Seed) &        -         &        -        & 	   0.218      &     -          & -0.167    \\
 \hline
      D &      0.187       &        -        & 	   0.124      &     -          & -0.167    \\
 \hline
      E &        -         &      0.099     & 	   0.119      &     -          & -0.016    \\
 \hline
      F &        -         &         -       & 	   0.121      &     0.194      & -0.018    \\
 \hline
   \end{tabular}
 \end{center}
 \caption{The WCs, as obtained from Fig.\ \ref{f:afby}, 
 for which Models A-F are separable from the 
 respective seed models at $3\sigma$ level. The integrated asymmetries are also shown.}
   \label{t:2cdafbxy} 
 \end{table}
 
 \begin{figure}[htbp]
\centering
  \subfloat[\label{RL-LL3}]
  {\includegraphics[height=5cm]{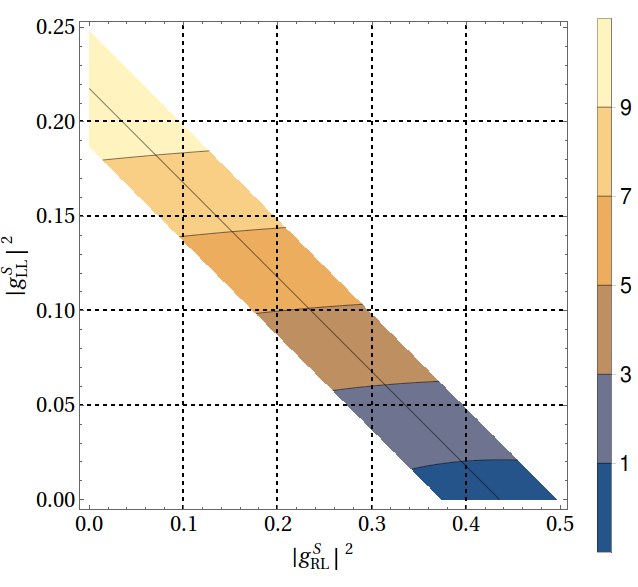}}
  \subfloat[\label{RL-RR3}]
  {\includegraphics[height=5cm]{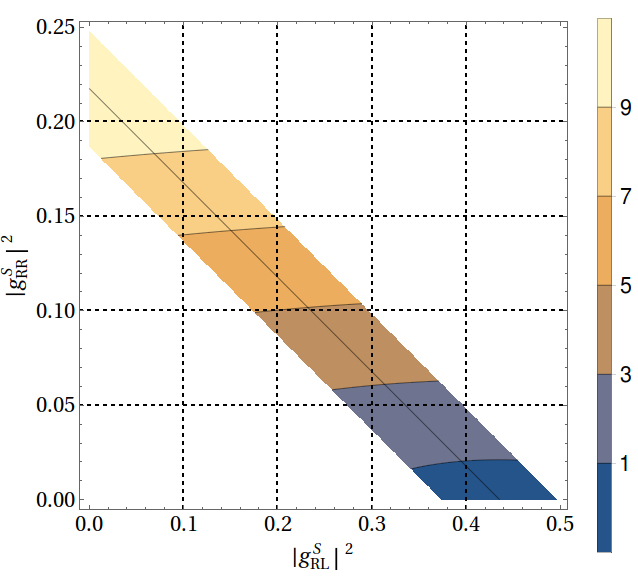}}
  \subfloat[\label{RL-LR3}]
  {\includegraphics[height=5cm]{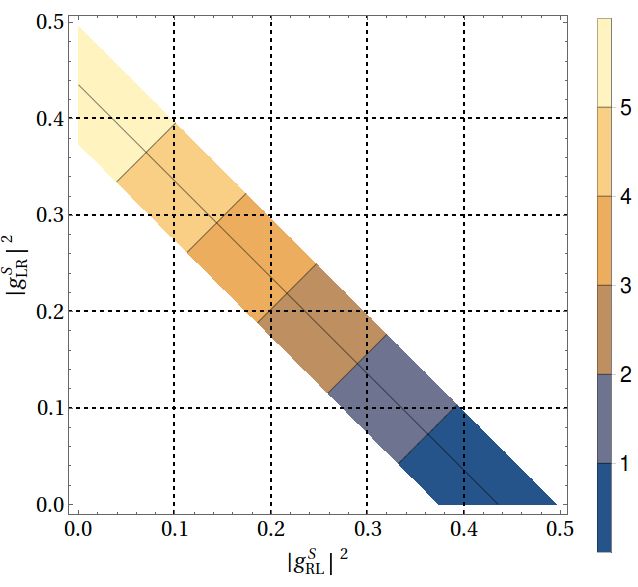}}\\
  \subfloat[\label{RR-RL3}]
  {\includegraphics[height=5cm]{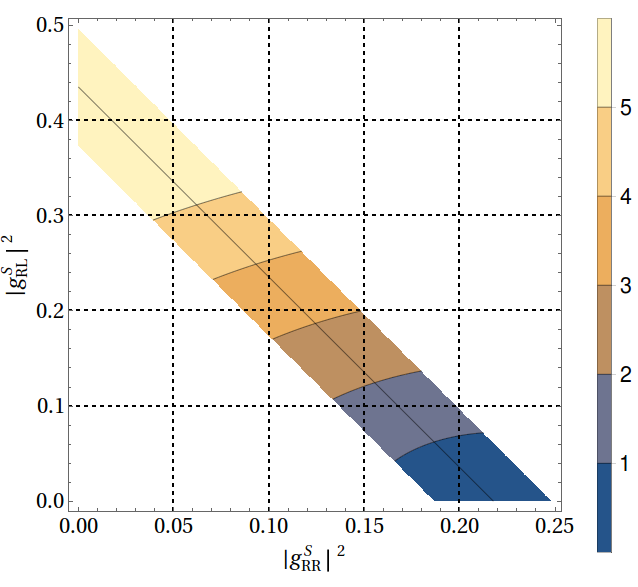}}
  \subfloat[\label{RR-LL3}]
  {\includegraphics[height=5cm]{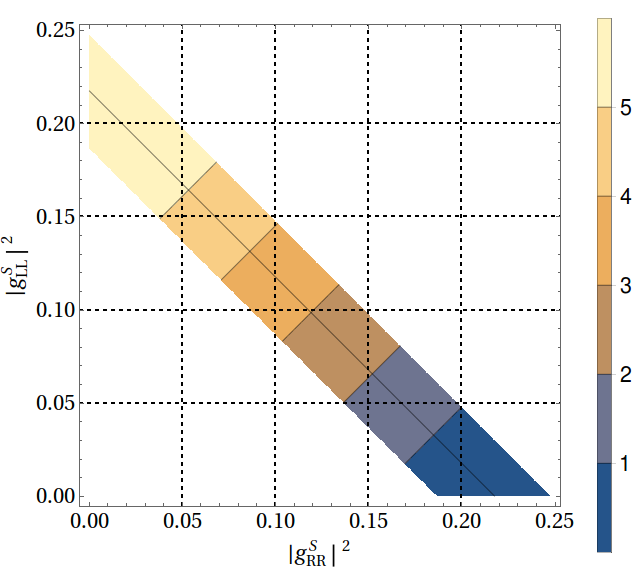}}
  \subfloat[\label{RR-LR3}]
  {\includegraphics[height=5cm]{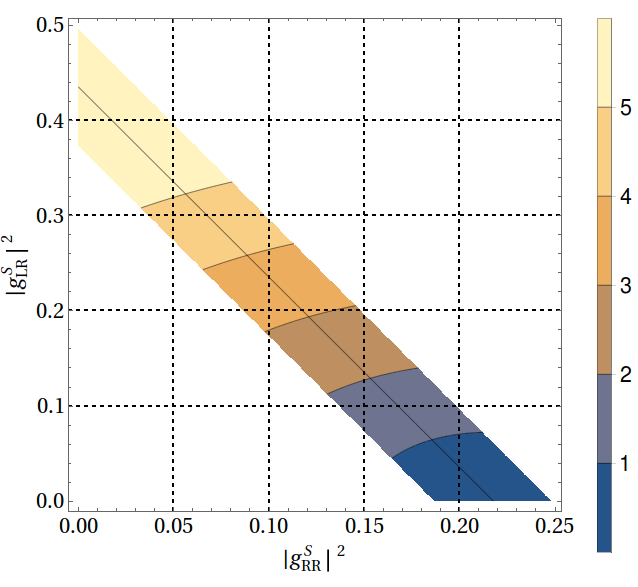}}
   
  \caption{The differentiability of the models A-F, shown in (a)-(f) respectively, from the 
  `seed' model, with ${\cal A}'_{FB}(y)$ as the observable.}
\label{f:afby}
\end{figure}

\begin{figure}[t]
\centering
  \subfloat[\label{2coupdist3}]
  {\includegraphics[height=5cm]{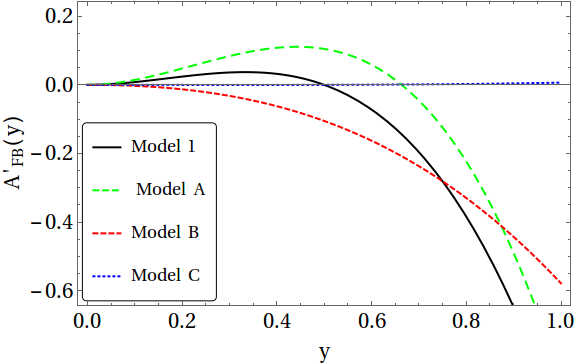}}
  \subfloat[\label{2coupdist4}]
  {\includegraphics[height=5cm]{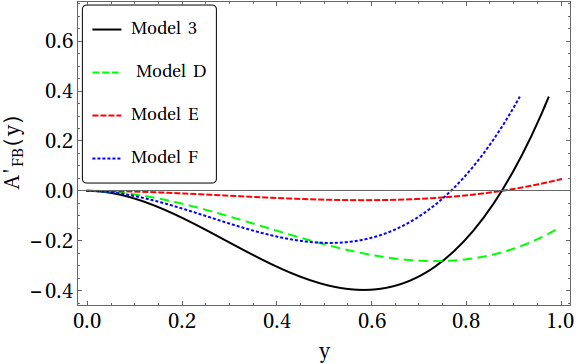}}
  \caption{Comparison of the distributions of the asymmetry ${\cal A}'_{FB}(y)$ for the models A-F from the 
  respective `seed' model. Here the relevant WCs are taken from the plots in figure \ref{f:afby}.}
\label{f:2cdafby}
\end{figure}

In Fig.\ \ref{f:2cdafby}, we show the distribution of ${\cal A}'_{FB}(y)$ for Models A-F, comparing 
A-C with Model 1 as seed and D-F with Model 3 as seed respectively. The corresponding WCs are given in 
Table \ref{t:2cdafbxy}. While a $3\sigma$ separation between the models is possible, one notes that
the differentiation works best in the middle-$y$ region, rather than at the endpoints.

\subsection{Observable: $dB_\tau/dx$ and $dB_\tau/dy$}

Study of the differential BRs is instructive. First, let us refer the reader to Tables \ref{t:2wcdbrdx} and 
\ref{t:2wcdbrdy} respectively for the coefficients $C_1$ and $C_2$ in all the models considered. For 
both $dB_\tau/dx$ and $dB_\tau/dy$, 
this shows immediately that Models A and B must yield identical distributions; same is true
for the pair D and F. This is because the BR does not depend on the change $R \leftrightarrow L$. 
Models C and E are very poorly differentiable from their respective seeds (at less than $1\sigma$) and so
we do not discuss them any further, neither do we show the corresponding separation plots. 

\begin{figure}[t]
\centering
  \subfloat[\label{RL-LL2}]
  {\includegraphics[height=3.9cm]{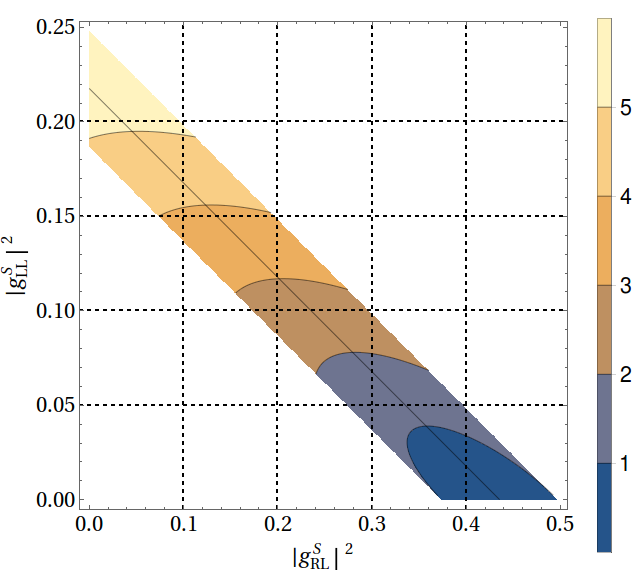}}
  \subfloat[\label{RR-RL2}]
  {\includegraphics[height=3.9cm]{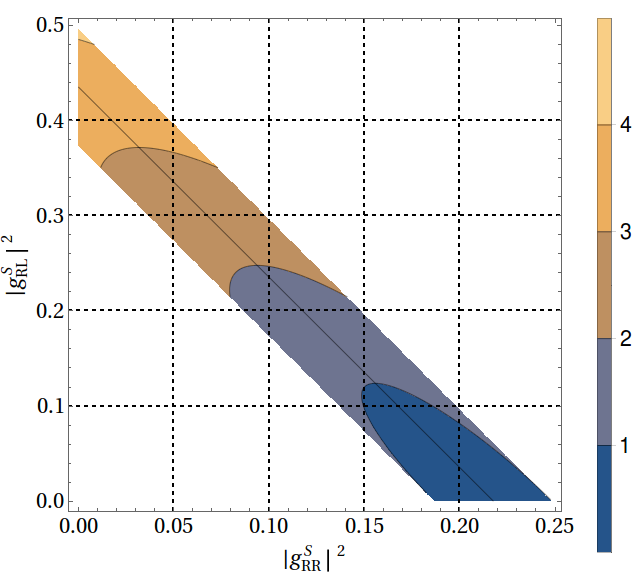}}
   \caption{The differentiability of the Models A and D, shown in (a) and (b) respectively, from the 
  `seed' model, with $dB_\tau/dx$ as the observable. Model B is identical with Model A, and Model F is 
  identical with Model D.
 }
\label{f:dbrdx}
\end{figure}

Even though the pattern seems similar, there is an important difference. With $dB_\tau/dx$ as the
observable, we can separate Models A(B) or D(F) from the corresponding seed models at $3\sigma$ 
or more, depending on the respective WCs. This can be seen from Figure \ref{f:dbrdx}, as well as Table 
\ref{t:2cddbx}. With $dB_\tau/dy$ as the observable, there is no available parameter space with 50
events where any model can be separated at more than $2\sigma$ from the seed models. This is why
we do not show the corresponding plots for $dB_\tau/dy$. Thus, as far as the measurement of the number
of events in different energy bins goes, it is preferable to detect the unlike-sign muon, than one of the 
like-sign muons. 

As the {\sf V} class of models show an identical behaviour, we conclude that based on only the data and 
without any {\em a priori} knowledge of the WCs, it is impossible to differentiate between the classes, 
but within a particular class, it is possible to differentiate among the various Lorentz structures of the
effective operators. 
With enough events, one should be able to differentiate single-operator models 
from the double-operator models, like those with pure $S$ or $V$ ($O_9$),  or with pure $P$ or
$A$ ($O_{10}$) muon current.
If we have approximately 50 events, $A'_{FB}(x)$ may help us differentiate 
$O_{9L}$ or $O_{10L}$ models from $RL$ by about $5\sigma$, and $O_{9R}$ and $O_{10R}$ models 
from $RR$ by about $7\sigma$. With $dB_\tau/dx$, 
the former set is differentiable at about $3\sigma$, while the latter is at less than $2\sigma$. (The $9(10)L
(R )$ models are specified by equal magnitudes of the two WCs.)
As $C_1$ and $C_2$ involve $\vert g_{IJ}\vert^2$, and hence it is insensitive to the sign or 
phase of the WCs.

\begin{table}[htbp]
 \begin{center}
 \begin{tabular}{|c|c|c|c|c|}
 \hline
 Model & Seed & Second operator & $C_1$ & $C_2$\\
 \hline
 A & $RL$ & $LL$ & $|g_{RL}^S|^2 + |g_{LL}^S|^2$ & $3 |g_{RL}^S|^2 + 2 |g_{LL}^S|^2$\\
 B & $RL$ & $RR$ & $|g_{RL}^S|^2 + |g_{RR}^S|^2$ & $3 |g_{RL}^S|^2 + 2 |g_{RR}^S|^2$\\
 C & $RL$ & $LR$ & $|g_{RL}^S|^2 + |g_{LR}^S|^2$ & $3 |g_{RL}^S|^2 + 3 |g_{LR}^S|^2$\\
 D & $RR$ & $RL$ & $ |g_{RR}^S|^2 +  |g_{RL}^S|^2$ & $2 |g_{RR}^S|^2 + 3 |g_{RL}^S|^2$\\
 E & $RR$ & $LL$ & $ |g_{RR}^S|^2 + |g_{LL}^S|^2$ & $ 2 |g_{RR}^S|^2 + 2 |g_{LL}^S|^2$\\
 F & $RR$ & $LR$ & $ |g_{RR}^S|^2 + |g_{LR}^S|^2$ & $2 |g_{RR}^S|^2 + 3 |g_{LR}^S|^2$\\
 \hline
  \end{tabular}
    \caption{$C_1$ and $C_2$ for different models. The observable is $dB_\tau/dx$.}
    \label{t:2wcdbrdx}
 \end{center}
 \end{table}

  \begin{table}[htbp]
 \begin{center}
 \begin{tabular}{|c|c|c|c|c|}
 \hline
 Model & Seed & Second operator & $C_1$ & $C_2$\\
\hline
 A & $RL$ & $LL$ & $ |g_{RL}^S|^2 +3 |g_{LL}^S|^2$ & $  |g_{RL}^S|^2 +4  |g_{LL}^S|^2$\\
 B & $RL$ & $RR$ & $|g_{RL}^S|^2 +3 |g_{RR}^S|^2$ & $ |g_{RL}^S|^2 +4 |g_{RR}^S|^2$\\
 C & $RL$ & $LR$ & $ |g_{RL}^S|^2 + |g_{LR}^S|^2$ & $|g_{RL}^S|^2 + |g_{LR}^S|^2$\\
 D & $RR$ & $RL$ & $ 3 |g_{RR}^S|^2 + |g_{RL}^S|^2$ & $ 4 |g_{RR}^S|^2 +  |g_{RL}^S|^2$\\
 E & $RR$ & $LL$ & $ 3 |g_{RR}^S|^2 + 3 |g_{LL}^S|^2$ & $ 4 |g_{RR}^S|^2 + 4 |g_{LL}^S|^2$\\
 F & $RR$ & $LR$ & $ 3 |g_{RR}^S|^2 +  |g_{LR}^S|^2$ & $ 4 |g_{RR}^S|^2 +  |g_{LR}^S|^2$\\
 \hline
 \end{tabular}   
     \caption{$C_1$ and $C_2$ for different models. The observable is $dB_\tau/dy$.}
     \label{t:2wcdbrdy}
 \end{center}
 \end{table}
 
\begin{table}[htbp]
 \begin{center}
\begin{tabular}{|c|c|c|c|c|}
  \hline  
 Model  &  $|g_{RL}^S|^2$  & $|g_{LL}^S|^2$  & $|g_{RR}^S|^2$ & $|g_{LR}^S|^2$  \\
  \hline
1(Seed) &      0.435       &      -          & 	   -          &     -           \\
 \hline
      A &      0.202       &     0.117       & 	   -          &     -           \\
  \hline
3(Seed) &        -         &        -        & 	   0.218      &     -           \\
 \hline
      D &      0.372      &        -        & 	   0.032      &     -           \\
 \hline
   \end{tabular}
\end{center}
 \caption{The representative values of the WCs, obtained from Fig.\ref{f:dbrdx}, for which Models A and D
 can be differentiated from the 
 respective seed models at $3\sigma$ level.}
  \label{t:2cddbx}
\end{table}   

In general, LFV models can also involve electrons in the final state, from operators leading to 
$\tau^-\to e^- e^+ e^-$ or $\tau^- \to e^- e^+ \mu^-$. The BRs of these channels have bounds comparable to 
that of $\tau\to 3\mu$, so we may expect a similar number of events at Belle-II, and a similar analysis 
will work. However, detection of final state electrons in an $e^-e^+$ machine will have lesser efficiency than 
that of final state muons. 

 \section{Conclusion}
 
In this paper, we focus on the LFV decay $\tau \to 3\mu$. This is of crucial importance in the light of 
semileptonic $B$-decay anomalies, which hint at some new physics involving second and third generation
leptons, probably a mixing among the charged leptons. The present limit on this mode translates to $\sim 70$
events at the most at Belle-II with 50 ab$^{-1}$ integrated luminosity. While even a single event will 
unequivocally indicate new physics, we try to answer a more ambitious question: is it possible to say 
anything about the underlying operators from the observables? Needless to say, the answer will be vital
for model builders. 

The answer to this question would have been much easier if final state muon polarisations could have been 
measured. As far as the present technologies go, this is not easily attainable. However, as we show, one can
form other observables, which are relatively clean and at the same time can yield significant information. 
One of the observables is the asymmetry of either the unlike-sign muon, or the like-sign more energetic 
muon, which is to be measured with respect to the initial $\tau$-polarisation direction. If one can measure
the asymmetries, even with the associated error margins, in different energy bins, this can differentiate 
between the different types of operators in a particular class (scalar, vector, or tensor). 

Another important observable, as expected, is the number of events in different energy bins, either the 
unlike-sign or the like-sign muons. Just like the asymmetries, it can
potentially differentiate among the different chiral structures of the operators, although to a lesser extent.
Given the total number of events, one can also have an idea of the magnitude of the relevant WCs. We
expect more events for {\sf V} or {\sf T} type operators, so their WCs, $g^V_{IJ}$ or $g^T_{IJ}$, can be 
probed better.

It may so happen that there are more than one NP operators. A typical case is when the muon current 
is purely vector or axial-vector in nature. If we have enough number of events ($\sim 50$), we should be able 
to say whether there is only one underlying operator or two. Asymmetries are the better observables,
but the distribution of the number of events can also help and act as complementary ones.

One must, however, remember that such an analysis involves the risk of underestimating the errors
by neglecting the systematic uncertainties. Thus, this is to be seen more as a motivation to the experimentalists.
Once the data is available, other powerful analysis methods, like the maximum likelihood, can be applied.

{\bf Acknowledgements}: 
AK thanks the Science and Engineering Research Board (SERB), Government
of India, for a research grant. SKP is supported by the grants IFA12-PH-34 and 
SERB/PHY/2016348.

\appendix


\end{document}